\documentclass{ws-jai}
\usepackage[flushleft]{threeparttable}
\begin{document}


\markboth{Baan, Jessner and Steenge}{Measuring data loss}

\title{Measuring Data Loss resulting from Interference}

\author{Willem A. Baan$^1$, Axel Jessner$^2$, Jaap Steenge$^3$}
\address{
$^{1}$Netherlands Institute for Radio Astronomy, ASTRON, Dwingeloo, The Netherlands, baan@astron.nl \\
$^{2}$Max Planck Institut f\"ur Radioastronomie, MPIfR, Bonn, Germany, jessner@mpifr.mpg.de \\
$^{3}$Radiocommunications Agency Netherlands, Ministry of Economic Affairs and Climate, Groningen, The Netherlands, \\ jaap.steenge@agentschaptelecom.nl}

\maketitle

\corres{$^{2}$Corresponding author Willem Baan.}

\begin{history}
\accepted{(Published in JAI Vol. 08, No. 01, 1940007 (2019))};
\end{history}

\begin{abstract}
This paper presents an observing methodology for calibrated measurements of radio interference levels and compare these with 
threshold interference limits that have been established for interference entering the bands allocated to the Radio Astronomy Service.
The measurement time and bandwidth intervals for these observations may be commensurate with the time and frequency variability characteristic of the interfering signals and the threshold levels may be appropriately scaled from the values presented 
in ITU-R RA.769 using a 2\,000 seconds reference time interval.
The data loss for astronomical instruments may be measured as a percentage of occupancy in the time-frequency domain 
both for short and long measurement intervals. 
The observed time-frequency occupancy characteristics for non-geostationary satellite systems and earth stations 
in the mobile-satellite service may be incorporated into an effective power flux density simulation to obtain the effective data loss and sky blockage due to these services.
 \end{abstract}


\section{Introduction}

Observations in radio astronomy aim to study the physical phenomena occurring in astronomical objects and to study known objects with a higher precision and sensitivity. The operational requirements for radio astronomy observations at the highest achievable sensitivity and spectral resolution require spfd (spectral power flux density) levels that are very much lower than those of stations in the active services. 

The percentage of data loss from interference is an important parameter for all radiocommunication systems. For radio astronomy specifically, any detrimental interference will reduce the quality of the data resulting in non-detection or alternatively a false detection of a radio source, or a reduction in the precision of radio astronomical measurements, perhaps even to the extent of making the measurement meaningless. Sometimes the lost information may be recovered by a repeat of the observation or by longer observations of the same object in the sky in the same frequency band. However, for measurements of singular events (e.g. supernova explosions, short flares or outbursts of compact objects, or cometary passages) the lost data cannot be recovered. 
In the context of this paper, we assume for simplicity that radio astronomical observations aim to measure a steady noise signal coming from an astronomical source within a specified bandwidth. We call the minimum integration time interval required to repeat a measurement that has been detrimentally affected by interference within a given band as the 'data loss' caused by the interference.

This paper presents the principles for a quantitative procedure of determining the percentage of data that is affected by interference above certain detrimental thresholds and refers to this as data loss. The actual impact of interference on astronomical data and the question whether some of this data may be partially usable is a qualitative matter that depends on the observing mode and the type of astronomical research project.  A rigorous quantitative assessment of the (qualitative) impact of interference on the data is beyond the intended scope of this document. 

This paper also describes the impact of interference on radio astronomy observations and the associated data loss in order to provide qualitative criteria for evaluating interference from active services operations in the shared, adjacent, nearby or harmonically related bands. 
Sections \ref{sec2} and \ref{sec3} discuss the criteria of data loss and present a quantitative methodology for measuring data loss. Section \ref{sec4} describes the determination of data loss based on epfd (equivalent power flux density) simulations. Section \ref{sec5} qualitatively describes the impact of interference on radio astronomical data. 

The Recommendations of the Radiocommunication Sector of the International Telecommunication Union (ITU-R) mentioned 
in this paper may be found at the ITU-R website as follows: 
for the Radio Astronomy Service \citep{ITU-RA}, 
for the Earth Exploration Service \citep{ITU-RS},
for the Mobile Satellite Service \citep{ITU-M}, and
for the Fixed Satellite Service \citep{ITU-S}.
Footnotes and the allocations of bands may be found in the ITU-R Radio Regulations \citep{ITU-RR}.

\section{Data loss in the radio astronomy service}\label{sec2}

The Recommendation ITU-R RA.1513 quantitatively defines data loss as the percentage of observations that contains interference above the detrimental threshold levels for continuum or spectral line observations as presented in Recommendation ITU-R RA.769. Further,  ITU-R RA.1513 determines a percentage data loss of 5\% to be used for the accumulated data loss caused by all systems in all services and all operators and a percentage data loss of 2\% to be used for a single system or network. 

The usefulness of radio astronomical data rapidly degrades in the presence of interference and astronomers are forced to discard the affected observations because the interference can produce imitation spectral line signals and/or raise the system noise level, resulting in incorrect measurements. The presence of interference contaminates the measurement data and impedes the scientific objectives of the experiment. 

Sensitive spectral line and continuum observations in selected bands allocated to the RAS on a primary basis require the prohibition of all emissions during the observations. Candidate bands for this condition are given in No. 5.340 of the Radio Regulations. The radio astronomical detector cannot in general discriminate between in-band emissions in a given passive band and unwanted emissions from a neighbouring band. Both cases will result in a degradation of the measurements; its severity depends solely on the characteristics of the interfering signal as measured in the RAS band.  In order to reduce the data loss in these particular bands to 0\%, the interfering signals should remain below the threshold levels of ITU-R RA.769 during any relevant time interval.

The ITU-R RA.1513 discusses a percentage of time for the data loss and does not stipulate the power levels of the interfering signals and their frequency behaviour. For the purpose of this recommendation, the percentage of data loss discussed in ITU-R RA.1513 has been interpreted to apply to the time-frequency occupancy of the interference.

The data loss criteria presented in ITU-R RA.1513 can also be incompatible with the detrimental threshold values of Recommendation ITU-R RA.769. While ITU R RA.1513 allows interference to surpass the equivalent threshold levels of ITU R RA.769 during 2\% or 5\% of time in primary allocated bands, the cumulative power of these signals may also exceed the threshold levels of ITU-R RA.769 for a 2\,000s reference time interval.  This then constitutes a 100\% data loss for the whole 2\,000s time interval.

For this reason, the criteria of both ITU-R RA.769 and ITU-R RA.1513 need to be applied in the evaluation of the percentages of data loss in RAS bands allocated on a primary basis. 

Data loss due to interference may result from loss of part or the entire observing band, part or all of the planned observing time, or from blockage of part of the sky. In the case of continuum observations, data loss may result when the integrated interference power in the RAS band in an observation is in excess of the detrimental threshold levels derived from ITU-R RA.769, independent of the frequency characteristics of the interference. In the case of spectral line observations, data loss may result if one or more channels inside the RAS band have interference power levels in excess of detrimental threshold levels for spectral line observations. Data loss for the RAS resulting from services in nearby bands may arise from three causes:

a)	Wanted and unwanted emissions falling in bands allocated to the RAS resulting from active radio services sharing the band with the radio astronomy service or from active systems in adjacent or nearby bands. 

b)	Intermodulation and departures from linearity in radio telescope systems may result from strong signals in adjacent and nearby bands. 

c)	Damage to radio telescope receivers from excessive power levels received (Annex 2 of ITU-R RA.1750).

The threshold levels of ITU-R RA.769 do not address the protection from interference received in the main beam or near sidelobes (main beam coupling) of the radio telescope, whereby a single interfering source can effectively remove several percent of the celestial sphere accessible for radio astronomical observations. Sky blockage may result from terrestrial transmitters and geo-stationary satellites (GSO) satellites as well as non-GSO satellites in lower-earth orbits passing through the primary beam and the inner-sidelobes of the radio astronomical antenna, as discussed in ITU-R RA.1513. Sky blockage caused by in-band (intended or unwanted) emissions in shared bands is classified as data loss.

Data loss caused by mobile systems and non-GSO satellite systems may be evaluated using simulations tools based on Monte Carlo models to determine the epfd (effective power flux density) generated by all transmissions of a particular active service. Three ITU-R Recommendations describe specific cases of data loss for the RAS. Recommendation ITU-R S.1586 describes unwanted emission levels produced by a non-geostationary fixed-satellite service system at radio astronomy sites while Recommendation ITU-R M.1583 describes interference calculations between non-geostationary mobile-satellite service or radio navigation-satellite service systems and radio astronomy telescope sites. Finally Recommendation ITU-R M.1316 describes the coordination between the RAS and mobile earth stations in the Mobile Satellite Service (MSS) (Earth-to-space) using a practical loss of 2\% observing time to determine the required {\it separation distance} between the distributed mobile earth stations in the MSS and the radio astronomy station. The Monte Carlo methodology used in these recommendations may also be extended to the evaluation of other terrestrial unwanted emissions into a radio astronomy band as well as new applications using space stations and high-altitude platform stations.

\section{Methodology to determine the percentage of data loss}{\label{sec3}

\subsection{The measuring and detection procedures}

The antenna to be used for measuring and detecting radio interference should be sensitive enough to be 'faster' than a radio astronomy antenna with 0 dBi sidelobe entry. This suggests that the forward gain of the antenna should be high enough and the system temperature of the receivers low enough in order to raise the sensitivity of the measuring system significantly over that of an isotropic receiver. The measuring antenna needs to be well calibrated in order to record spfd and pfd values at the levels that are typical for radio astronomy instrumentation and as presented in Recommendation ITU-R RA.769. Since the threshold values in Recommendation ITU-R RA.769 are determined for 0 dBi sidelobe entry and not for main beam entry, a measuring system with a significant forward gain that is able to point at the interfering sources will achieve the detection of interfering signals at levels close to the threshold values of Recommendation ITU-R RA.769 within shorter measurement intervals than an omni-directional system. An omni-directional measuring antenna would be less effective for this purpose. The measuring antenna should also be able to point at the horizon for terrestrial sources and be able to track fast moving targets in the sky. Because most astronomical antennas cannot easily track rapidly moving sources or point at the horizon, it is not customary or even feasible to use an astronomical antenna for these purposes.

The standard detection procedure used at single dish radio telescopes requires a drift scan for continuum signal or ON-OFF sequences for both continuum and spectral line measurements. This sequence requires an observation at the ON-source position and an equally long OFF-source observation on a position of blank sky close to the source position. In practise for the purpose of RFI detections, the OFF-source observation can take place after a much longer ON-source measurement and be taken at a nearby antenna position (for instance at same elevation and different azimuth position), as long as the reference observation represents a RFI-free environment, i.e. a 'blank' sky equivalent.  The RFI signature is then determined by the (ON minus-OFF) signal taking into account the difference in duration for the ON and OFF. For astronomical measurements the (ON-minus-OFF)/OFF ratio is used to also calibrate the bandpass shape. 
The measuring system should adopt the same data collecting procedures as astronomical observatories, where observations are routinely done using short (typically 1 second or less) measurement intervals. During processing the individual data records may be averaged (stacked) in order to achieve higher sensitivity for continuum and spectral line detections.

The system parameters of the measuring station are determined by the regimes of power flux density that are expected for the interference to be measured. The forward gain of the measuring telescope needs to be at least 30 dB in order to detect signals that are within the noise of the RAS system but above the ITU-R RA.769 detrimental threshold. The measuring system may have spectral channel widths that are different from the {\it reference spectral bandwidths} presented in ITU-R RA.769, which range from 10 kHz below 1 GHz to 1 MHz above 60 MHz, as long as the proper bandwidth conversion is done during the analysis.  The sampling time interval of the system should be as low as one second and may have to be reduced further to a fraction of 1 sec in special cases. Length of the measurement series should be at least 1\,000 seconds or as long as possible in the case of a satellite passage.  The measurements for the determination of data loss are done as a time series of N measurements with a time interval $\Delta t$ covering M frequency channels with bandwidth $\Delta f$.

The measuring telescope should avoid pointing at the horizon for the case of satellite monitoring in order to avoid confusion with terrestrial sources. On the other hand, pointing at the horizon is required for terrestrial sources of interference and for a complex terrestrial environment the spectral signature of each interfering source should be identified separately.  In addition, adequate filtering should be applied to avoid any non-linearity issues in the receiver and the generation of internal intermodulation and to measure weak out-of-band emissions generated by strong external interfering signals.  

\subsection{The signal levels and thresholds}\label{sec3.2}

The signals from cosmic radio sources received by radio astronomy stations are usually hidden within the noise of the individual observation and can only be recognised after summing multiple observations. Because of white noise characteristics for the receiver system and the sky, the use of larger bandwidths and/or longer integrations reduces the rms noise level $\Delta T$ for an astronomical observation as: 
\begin{equation}
		\Delta T =  T_{sys} / \sqrt (\Delta t\Delta f),	
\end{equation}
\noindent where $T_{sys}$ is the system temperature of the astronomical receiver system and $\Delta t$ is the time interval of the observation. The observing bandwidth $\Delta f$ is the channel width for high-spectral-resolution spectral line observations and the allocated bandwidth (or wider) for the case of broadband continuum observations. The designated values of $T_{sys}$ for the astronomical receiver system reflect current technology practises and range from 22K to 100K depending on the observing frequency as indicated in ITU-R RA.769. By employing antennas with gains of 50-80 dBi, wide observing bandwidths, and long integration times (using data averaging), and (further) reducing system noise temperature of the receivers, the astronomers routinely measure celestial flux densities ~60 to 90 dB below the harmful levels defined in ITU-R RA.769.  

A quantitative evaluation of the percentage data loss in RAS bands is based on the identification of interference signals in the data records that are in excess of the levels of detrimental interference presented in Recommendation ITU-R RA.769, which are based on a well-defined {\it reference time interval} ($\Delta t$) of 2,000 seconds (33 minutes) and a well-defined reference channel width ($\Delta f_{ref}$) or reference continuum bandwidth ($\Delta f_b$). 

The detrimental threshold flux density levels $T_{spec}$ and $T_{cont}$ to be used during the evaluation of data loss in RAS bands may be derived from the spfd values for broadband continuum (subscript cont) and narrow band spectral line (subscript spec) observations in  ITU-R RA.769 and corrected for the bandwidth and duration used for the measurement: 
\begin{equation}
T_{spec}(\Delta t,\Delta f)  =  spfd_{spec}(RA.769, table 2)  + 5 log ((\Delta f/\Delta f_{ref}) (2000s/\Delta t))  (dB(W/m^2/Hz)), 
\end{equation}
\begin{equation}
T_{cont}(\Delta t, \Delta f)  =  spfd_{cont}(RA.769, table 1)  + 5 log ((\Delta f/\Delta f_{b}) (2000s/\Delta t))  (dB(W/m^2/Hz)),
\end{equation}
\noindent where 2\,000s is the reference time interval. For spectral line observations the value to be used for $\Delta f_{ref}$ equals the channel bandwidth given in ITU-R RA.769, while $\Delta f$ may depend on the characteristics of the detection system used. For continuum observations the reference bandwidth $\Delta f_b$ should equal the allocated RAS bandwidth. 

The detection levels of the measuring antenna will have a similar dependence on the measurement interval and spectral bandwidth as described in the above equations. The detection thresholds need to be corrected for the gain of the antenna and the system temperature of the measuring system as the ratio of the two temperatures (see Equation 1). The threshold levels of ITU R RA.769 for the astronomical antenna system are based on signals that are 10\% of the integrated noise fluctuations in the detection system. The actual detection (1 sigma) thresholds of the measuring antenna will also be 10 dB higher. The detection levels for main beam entry with the measuring antenna may thus be expressed as:
\begin{equation}
S_{spec}(\Delta t, \Delta f)   = -(G+10) + [10 log (T_{sys,ref/}T_{sys,mon)}] + T_{spec}(\Delta t, \Delta f)  (dB(W/m^2/Hz)), 
\end{equation}
\begin{equation}
S_{cont}(\Delta t, \Delta f)     = -(G+10) + [10 log (T_{sys,ref}/T_{sys,mon})] + T_{cont}(\Delta t, \Delta f)  (dB(W/m^2/Hz)),
\end{equation}

\noindent where the levels $T_{spec}(\Delta t, \Delta f)$ and $T_{cont}(\Delta t, \Delta f)$  from RA.769 have been derived in Equations (2) and (3), and the forward gain $G$ of the measuring antenna is expressed in dB. The terms in Equations (4 and 5) containing the system temperature of the measuring system corrects for the difference between the system temperature quoted in ITU-R RA.769 ($T_{sys,ref}$) and the system temperature of the measuring system ($T_{sys,mon}$).

If the measuring receiver system has the same sensitivity as  the detection systems used at radio astronomy stations, then the measurements of interfering signals that are at (or close to) the actual levels of ITU-R RA.769 will require integration periods that are at least 100 times longer than the measurement interval. The forward gain of the measuring antenna reduces the time interval required for detection.  However, as the actual detection levels for shorter measurement intervals ($\Delta t$) increase as [5 log (2000s/$\Delta t$)] dB (see Eq. 1), only increasingly stronger interference signals can be detected for shorter time intervals at the measuring system. 

\subsection{Phase A -- Calibration of the system}

Assuming that a small aperture telescope with forward gain $G$ is used for detection and measuring of the interference, this telescope needs to be well calibrated. In particular, the conversion from antenna temperature ($T_A$) (or equivalently from the recorded noise power $P$) to spectral power flux density units of spfd (dB(W/m$^2$/Hz)) or Jansky (1 Jy = 10$^{-26}$ W/m$^2$/Hz or -260 dB(W/m$^2$/Hz)) should be well known as a function of frequency. In order to establish this calibration conversion, the following procedure may be used in preparation for upcoming interference measurements: 

1) 	Perform an ON observation of a few strong celestial continuum sources with known flux density (in Jansky) at the right frequency, such as Centaurus A, Virgo A, Cassiopeia A, Cygnus A, or other strong calibrator sources, and 

2) 	Perform an OFF observation for each source of a nearby piece of 'blank' sky for the determination of the zero signal baseline. 
The source strength at the given frequency divided by the measured signal difference
($T_A$(ON) - $T_A$(OFF)) provides the conversion factor between antenna temperature (in temperature units K) and flux units (in Jansky or as pdf in dB(W/m$^{-2}$/Hz)) for each source as a function of frequency. This conversion parameter is often called the {\it sensitivity} and is expressed in units of (Jy/K). 

The conversion parameter obtained in this manner can subsequently be used for converting the ON-source signal strength of detected interference signals from antenna temperature $T_A$ to spfd units and Jansky.
Calibrating the interference signals can also be achieved by switching between the antenna and a load (a Dickey switch) that has previously been calibrated using celestial sources.

\begin{figure}
\begin{center}
\includegraphics[width= 9cm]{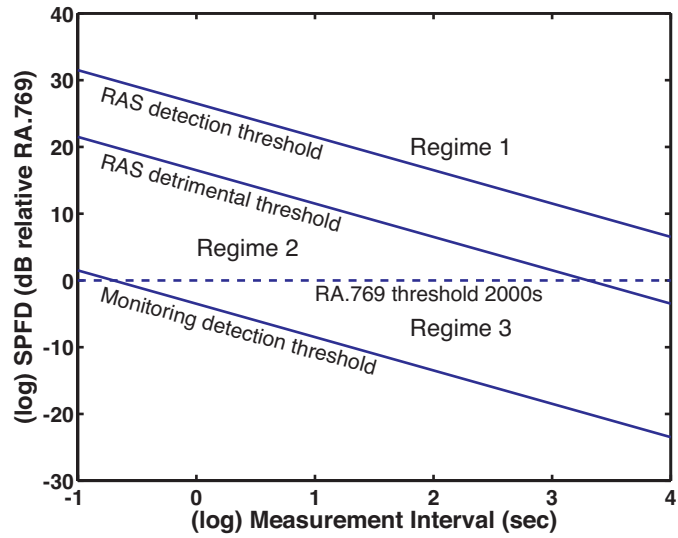}
\end{center}
\caption{A diagram showing the scaling of power flux density levels of interference signals compared with the values of ITU-R RA.769 as a function of the measurement interval. The line labelled as {\it RAS detrimental threshold} is obtained from Equation (2, 3) using the scaling of the measurement interval and is 10 dB below the (0 dBi) {\it RAS detection threshold}. For the purpose of illustration, the RFI Measuring system is assumed to have a forward gain $G$ = 30 dB, which lowers the detection threshold of the system. Three interference regimes may be identified: a) interference in Regime 1 is always above the detrimental level of ITU-R RA.769 for any measurement interval and account for the percentage data loss, b) interference in Regime 2 lies below the instantaneous RAS Detrimental threshold and may also add to the percentage data loss. The signal will reach above ITU-R RA.769 thresholds if the signal is continuous or has high enough time occupancy, and c) interference in Regime 3 is always below the RA.769 threshold.}
\label{fig:fig1}
\end{figure}

\subsection{Phase B -- Identification of the RFI}

A source causing detrimental interference may be identified using a single ON-source measurement with a longer time interval of T seconds and a measurement bandwidth $\Delta f$. A comparison OFF source measurement may be made at a position away from the source of interference. In the case of spectral line use of the allocated RAS band, all spectral channels of bandwidth $\Delta f$ located within the RAS band should be evaluated independently for the presence of interference above the spectral line detrimental thresholds using the Equation (2). In the case of continuum use of the allocated RAS band, the entire band should be evaluated using Equation (3). 

The levels of interference measured with a RFI Measuring (RFIM) system with a forward gain $G$ may be depicted relative to the detrimental values from RA.769 for different values of the measurement interval in Figure \ref{fig:fig1}. The detrimental threshold levels of Equation (2, 3) for spectral line and continuum measurements are 10 dB below the experimental detection levels of the RAS system (i.e. they lie within the noise floor) for an omni-directional system with 0 dB gain or a radio telescope with 0 dBi sidelobe entry. The detection noise floor for the RFIM system will have the same dependence on the time interval but will be lowered by the value of the forward gain, assuming the system temperature of the measurement system is similar to that of the radio astronomy antenna. The measured interference signals may thus be qualified in three Measurement Regimes as shown in Figure \ref{fig:fig1}: 

{\bf Measurement Regime 1} -- Interfering signals that are measured to be in Regime 1 lying above the line of the RAS Detrimental Threshold will qualify as detrimental interference for any used measurement interval and will add up to the percentage of data loss.  Integrated over 2000 seconds such signals will also exceed the thresholds given for continuum and spectral line measurements in ITU-R RA.769.  Interference signals in this Measurement Regime will be easily detected in the observational data and are most damaging to the data quality.

{\bf Measurement Regime 2} -- Interfering signals in Regime 2 lie below the RAS Detrimental Thresholds and above the RA.769 thresholds for the shorter time intervals. These signals may still be qualified as detrimental interference if the signal is continuous or has a high enough time occupancy. For instance, a signal at 10 dB above ITU-R RA.769 in a two second time interval with duty cycle $P$ = 10\% will be just at the threshold for a 2\,000 second time interval. 
Interference signals in Regime 2 will only be recognised after integration of the data and may also add to the percentage data loss.

{\bf Measurement Regime 3} -- Signals in Regime 3 below the RA.769 thresholds will not qualify as detrimental interference.

\subsection{Phase C -- Measurement of data loss using shorter time intervals}

The presence of sporadic, time-variable, and also time-invariable interference, may be determined using shorter measurement time intervals. A series of $N$ observations ON-source with a measurement interval $\Delta t$ may be evaluated for an intended observing period of $N \Delta t$, where $\Delta t$ must be appropriately small and $N$ sufficiently large to fully describe the characteristics of the interfering signals. A comparison OFF-source measurement may be made at a position away from the source of interference, which may be used for calibration of the bandpass of the detector system.

After calibration of the data records and conversion from antenna temperature (K) to flux density units, the data record for each measurement interval should be evaluated for the presence of interference that exceeds the corresponding (corrected) detrimental threshold levels for spectral line or continuum (see Eq. 2, 3). This procedure produces a time-frequency occupancy diagram of the detrimental interference in the measurements (waterfall plot) with dimensions of $N$ records and $M$ spectral channels.

\begin{figure}
\begin{center}
\includegraphics[width= 9cm]{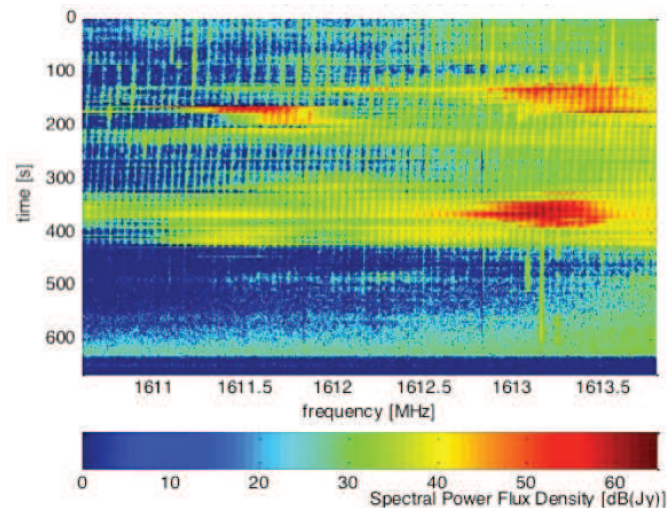}
\end{center}
\caption{Time-frequency occupancy diagram for the unwanted emissions emitted in the RAS band 1\,610.6 --1\,613.8 MHz during a complete passage of a single LEO satellite operating in an adjacent band. The measurement interval is 1 second and the measurement channel bandwidth is 6.1 kHz, which gives a threshold level $T_{spec}$ = 36 dB(Jy) (see Equation (2)). This diagram with $N$ = 630 data records and $M$ = 420 spectral channels covering the RAS band shows that interference can be detected both below and above the detrimental threshold levels of RA.769. The percentage of data loss varies from about 10\% at low frequencies to 60\% at the highest frequencies. This data has been obtained at the Satellite-Monitoring Station in Leeheim operated by the German Federal Network Agency (Bundesnetzagentur) using a spectrometer of the Max Planck Institute for Radio Astronomy. Example taken from ECC Report 171.}
\label{fig:fig2}
\end{figure}

In the case of continuum observations, the used bandwidth ($M\Delta f$) should cover part or the entire allocated RAS band. In the case of spectral line observations, all spectral channels of bandwidth $\Delta f$ located within the RAS band should be evaluated separately for the presence of interference above the spectral line detrimental thresholds $T_{spec}(\Delta t, \Delta f)$ (Eq. 2).  

The choice of the measurement time interval is important and should be governed by the time characteristics of the interference to be measured. The use of shorter intervals will help to isolate individual interference event but also raises the detrimental threshold and the detection threshold of the measuring system. Larger intervals will smear interference events and may raise the percentage data loss. 
For persistent or stationary (time-invariable) interference, it is sufficient to establish the presence of detrimental interference in a short time interval that fulfils the criteria of Equations (2, 3). For such interference, the number N of required measurements might be limited. 
Interfering systems with a fixed duty cycle emit power during a fraction of this cycle. Because radio astronomy systems cannot routinely record data on time-scales of less than one second, the actual data loss caused by interferers with small duty cycles may be much larger than suggested by the actual measurement of this duty cycle. 

Spread spectrum interferers emit low power levels across a larger fraction of the bandwidth with high duty cycles. ON-OFF measurements of spread spectrum interference need to cover sufficient bandwidth in order to detect the spfd enhancement.
In general, the measurement time interval should be short enough to be able to reveal the time variability and intermittence of the interfering signal. However, when the interferer is persistent and using frequency variability or frequency sweeping, the use of a short time interval may put the signals below the detection threshold. In this case an integrated measurement with a time interval covering the period of the variability or the sweep cycle is more appropriate (see Section \ref{sec5} below).

The measurement methodology described in these sections has been followed to obtain calibrated data of the rapidly varying unwanted (out-of-band) emissions in the RAS band 1\,610.6 --1\,613.8 MHz caused by a single LEO satellite operating in an adjacent band.  The 12m Satellite-Monitoring Station in Leeheim operated by the German Federal Network Agency (Bundesnetzagentur) with a a forward gain $G$ = 44 dBi (1.5 - 1.8 GHz) and a $T_{sys}$(zenith) = 120 K has been used to track individual LEO satellites. Using a spectrometer of the Max Planck Institute for Radio Astronomy, a calibrated time-frequency diagram was obtained with spfd detection levels well below the RAS detrimental threshold $T_{spec}$ = 36 dB(Jy) for 1 second measurement intervals (see Equation (2)).
The time-frequency diagram presented in Figure \ref{fig:fig2} with $N$ = 630 data records and $M$ = 420 spectral channels covering the RAS band and detected during one satellite passage shows that interference can be detected both below and above the detrimental threshold levels of ITU-R RA.769. The percentage of data loss varies from about 10\% at low frequencies to 60\% at the highest frequencies. Details of these measurements have been presented in ECC Reports 171 and 247.

\subsection{Determining the percentage of data loss}

The percentage data loss in a series of N measurements can be determined from the determination of the number of observations P that have interference levels in excess of the (corrected) spectral line $T_{spec}(\Delta t, \Delta f)$ or continuum $T_{cont}(\Delta t, \Delta f)$ threshold levels.  In practise this means counting the pixels in a time-frequency diagram with elevated flux density levels.

{\it \bf Time Series analysis:} The assessment of data loss for continuum observations requires that the aggregate flux density in all frequency channels of the RAS band be evaluated for excess flux density, which condenses a time-occupancy diagram into a simple time series. Similarly a time series analysis for spectral line data in a time-occupancy diagram requires that each frequency channel be evaluated for each time frame (each data record).  If one channel in a spectral line data record exceeds the threshold the whole record gets flagged.

If $P$ out of $N$ observational data records exhibit interference in one or more channels above the spectral line threshold values or in the whole RAS bandwidth above the continuum threshold, to first order the percentage data loss would be $P/N$ x 100\%, which is valid for large $N$. A large number of independent observations N is required to achieve a high ($\geq$ 95\%) confidence level for the actual percentage of data loss, particularly for small values of $P/N$. In Figure 2 the percentage data loss has been determined in this manner to be about 10\% at lower frequencies and 60\% at higher frequencies. 

For the case that the occurrence of interference (i.e. its arrival within the measurement series) is unpredictable, the (limited) number of the measurements $N$ needs to be taken into account and the percentage of data loss is expressed as $(P+1)/(N+2)$ x 100\% using small number statistics. This expresses the probability that interference randomly distributed in time can produce interference in $P$ out of $N$ measurements. 

For the case that the interferer has a regular (time-invariable) duty cycle or repetition rate, the percentage of data loss based on a representative measurement would again be $P/N$ x 100\%, which also assumes that the number of measurements $N$ is large.

{\it \bf Time-occupancy analysis:}  Analysis of the time-occupancy diagram for time and frequency variable interference requires that the power flux density in each ($\Delta t, \Delta f$) element of the time-frequency diagram needs to be evaluated individually for excess flux density.  Alternatively to a time series (sequential) evaluation, it is also possible to count the number of pixels with excess flux density. Assuming that the analysis shows that $P$ pixels from the total $N*M$ pixels within the RAS part of the time-occupancy diagram have flux density values above the corrected spfd threshold value, the data loss simply follows from the ratio $P/(N*M)$ x 100\% for this observation.  Alternatively, the frequency axis may be regridded before analysis in order to be conforming to the $\Delta f_{ref}$ value used for the band in Recommendation ITU-R RA.769. With coarser frequency gridding, the resulting percentage data loss will likely go up. Evaluation of the time-frequency diagram of Figure \ref{fig:fig2} in this manner results in a percentage data loss of about 40\%.

\subsection{Phase D -- Measurement of data loss using longer time intervals }

In addition to using short measurement time intervals for determining the percentage of data loss, a measurement of the accumulated flux density during longer intervals may serve as a second quality criterion for data loss. Rather than emphasising the short-term variability, the accumulated flux density determines the long-term behaviour of the signals during a measurement time interval $N \Delta t$. The use of longer time intervals to determine data loss will lower the detrimental threshold levels but will also designate larger blocks of data as data loss.  Interference signals that exhibit an acceptable percentage of data loss in the Regimes 1 and 2 of Figure 1 using a measurement interval $\Delta t$ may still exceed the threshold levels for a longer interval $N \Delta t$.

The integrated spectrum may be obtained by stacking the N data records after calibration, bandpass correction, and conversion to flux density units as described in Phases A-C above. The interference in the integrated spectrum may be evaluated using the threshold levels of Recommendation ITU-R RA.769 adjusted for a time interval of $N \Delta t$ seconds, as presented in Equations (2, 3) and (4, 5) of Section \ref{sec3.2} above. Data that contains interference exceeding these threshold levels in the selected bands may also be evaluated as data loss. Recommendation ITU R RA.769 presents threshold values for such a longer-term evaluation using a reference time interval of 2000 seconds.This procedure may be applicable for selected primary bands for which the time-integrated interference should not exceed the detrimental thresholds of Recommendation ITU-R RA.769 for continuum and spectral line observations, such as the primary bands mentioned in No. 5.340.

\section{Determining data loss from epfd simulations} \label{sec4}

Monte Carlo methods and alternative determinations may be used to determine the effective pfd levels for distributed interfering systems. Recommendations have been developed to address the specific methodologies for evaluating data loss from non-geostationary fixed-satellite systems (ITU-R S.1586), non-geostationary mobile-satellite or radio navigation-satellite systems (ITU-R S.1583), or mobile stations in the MSS (Earth-to-space;  ITU-R M.1316). While these simulation tools do not involve the actual detection levels, they should take account for the Doppler effect and the pfd differences from one interferer to the other. 

The input information required for epfd simulations are measurement data acquired using the methodology described above in order to describe the data loss resulting from interference from operational systems. Alternatively simulated/calculated emission characteristics may be used to determine the data loss resulting from future or operational system. 
The epfd simulation algorithm flow requires two iteration loops as summarised below:

{\it For each trial and for each cell on the sky the following steps are required:} a) Determine a random pointing of the telescope within a cell on the sky, and b) Determine a random initial time of the simulation $T_0$.

{\it For each interferer the following steps are required:} a) Calculate the position of the interferer for each second of the integration time, starting at $T_0$;  b) Determine for each time step a pfd value following the observed spectral distribution; c) Calculate the RAS antenna gain in the direction of the interferer, and d) Calculate the sum of the products of the RAS antenna gain and the pfd.

The epfd at the RAS station may be obtained by linearly summing the contribution of each interferer, then divide by the maximum antenna gain to get an epfd instead of an aggregate pfd, then determine the mean value over the integration time.
Finally the epfd value in dB should be compared to the appropriate threshold value given in ITU-R RA.769 minus the maximum antenna gain to determine whether there is a data loss.

These trials are to be repeated to obtain a statistical significance; for instance, 100 trials to determine the epfd ensemble average per cell are required to give a resolution of 10\%, because the accuracy of the simulation only improves as the root of the number of trials per cell.  If all 2334 (1$^o$ x 1$^o$) sky cells have the same statistical distribution around their individual mean, which generally they do not, then to first order the average data loss figure will have an accuracy of 0.021 times the standard deviation per cell. One hundred trials would be just sufficient to determine a data loss of 2\% $\pm$0.2 \%.  Data loss as a percentage of trials can be calculated for each cell, and the average can be calculated for the whole sky as the average of all cells.

This methodology may be used to independently determine the cumulative data loss and also the sky blockage resulting from emissions of a non-geostationary mobile-satellite system using measurements of multiple passes of single satellites, such as for the single passage shown in Figure \ref{fig:fig2}. This epfd MC methodology has been applied to determine the data loss in the Primary RAS band 1\,610.6 --1\,613.8 MHz caused by the Iridium system and has been detailed in ECC Reports 171 and 247. Examples of the resulting sky blockage are presented in Figure \ref{fig:fig3}.  The percentage of data loss diagram for a 30\,s (left) integration shows that substantial data loss happens close to the horizon at the begin and the end of a satellite passage where satellite antenna pattern points at the measuring antenna. The diagram for a 2000\,s (right) data integration suggests almost 100\% data loss across the sky. 

\begin{figure}
\begin{center}
\includegraphics[width= 8.5cm]{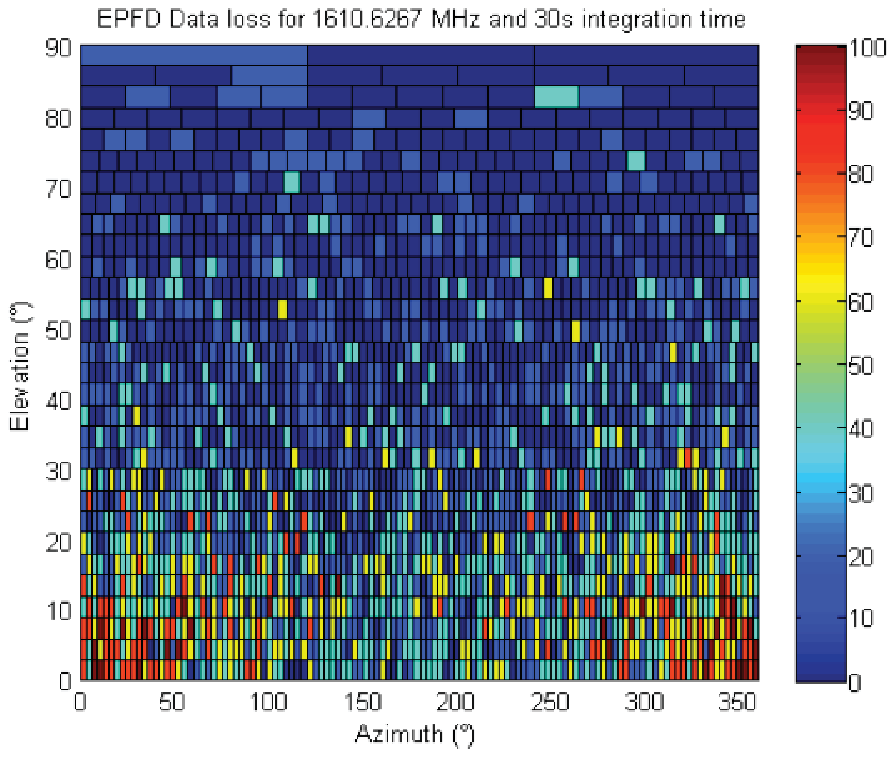}
\includegraphics[width= 8.5cm]{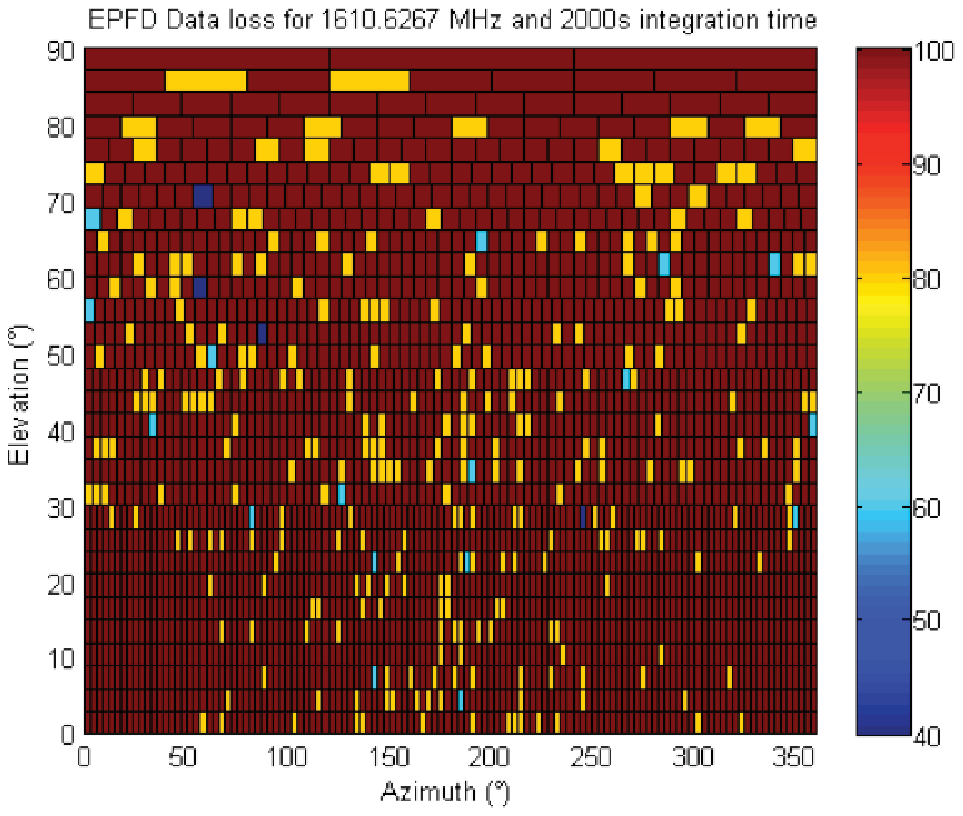}
\end{center}
\caption{Distribution of the EPFD data loss across the sky resulting from unwanted emissions in the RAS band 1\,610.6 --1\,613.8 MHz caused by the Iridium LEO Mobile Satellite system operating in an adjacent band.  These simulated result shave been obtained using the Monte Carlo methodology described in Section \ref{sec4} using data for single satellite passes such as presented in Figure \ref{fig:fig2}. The percentage of data loss diagram for a 30\,s (left) integration shows that substantial data loss happens close to the horizon at the begin and the end of a satellite passage where satellite antenna pattern points at the measuring antenna. The diagram for a 2000\,s (right) data integration suggests almost 100\% data loss across the sky.  Examples taken from ECC Report 247.}
\label{fig:fig3}
\end{figure}

\section{The impact of data loss on the RAS observations}\label{sec5}

The evaluation of the impact of the interference on radio astronomical observations is a complex issue that cannot easily be quantified and qualified in a general manner. In general, the impact of interference is greatest for single-dish instruments and decreases for multiple telescope interferometry systems using increasingly longer baselines. Localised interference will survive in the data during an auto-correlation process but will increasingly be suppressed during cross-correlation for the longer baselines of array instruments \citep{Thompson1982, FridmanB2001}. Multiplying interferometers and more complex synthesis arrays discriminate against interfering signals because of the relative phase changes of the signals received in distributed antennas, associated with the sidereal motion of a cosmic source across the sky. Signals that do not show this predictable phase behaviour are substantially suppressed in the data processing. 
The major uncertainty for both single-dish or array instruments results from the variability of the gain of the antenna side-lobes in which the interference is received. 
Parameters that influence the impact are the strength of the interference, its frequency and time characteristics, the location of the interference within the spectrum for the case of spectral line observations, the type of observations, and the known or expected characteristics of the astronomical signal. As a result, the methodology presented above does not address the qualitative aspects of the impact of the measured data loss. 

Generally, the astronomical signal is hidden within the noise of the individual observations and can only be recognised after summing multiple observations. In theory the degradation of quality of astronomical data may be compensated by a longer integration period if the interference signals also were to have white noise characteristics, a theoretical concept that is approximated only in nature. In reality, man-made interfering signals do not show the statistics of random noise, and longer integration times do not lead to a reduction of the strength of interfering signals. Moreover, if the interfering signal becomes comparable to or greater than the normal thermal fluctuations of the integrated receiver output, the necessary additional observing time becomes excessive and a precise estimation of the uncertainty in the measurement is no longer possible. As a result, spectral contamination rapidly brings a dominant portion of the data quality outside the observer's control.  

It should be noted that the allocations to the RAS in the  Table of Frequency Allocations in Article 5 of the Radio Regulations have certain designations for their use as for continuum and spectral line observations. Some allocations are too narrow to be used for continuum observations and other bands do not contain any observable spectral lines. Recommendation ITU-R RA.769 and the Radio Astronomy Handbook describe the use for each of the allocated RAS bands. 

In the following sections the impact of interference on RAS observations will be described and the way the astronomer can manage this impact. While the RAS stations employ larger operational bandwidth, these discussions only apply to the RAS bands allocated in the Table of Frequency Allocations of the  Radio Regulations.

\subsection{Continuum observations} 

Continuum observations employ the whole allocated radio astronomy band as a reference bandwidth. Actual observations may also be done by dividing the band into narrow channels and integrating (summing) those channels during subsequent signal processing. The quantitative time series determination of the percentage data loss from Section 3 applies for continuum measurements.

{\it \bf Broadband interference} --
The effect of broadband interference in excess of the detrimental thresholds given in Recommendation ITU-R RA.769 for continuum observations results in a complete loss of all or part of the data records over the radio astronomy band. Data loss will be measured as a percentage of time using the time series of integrated data records. 

{\it \bf Narrowband interference} --
Narrowband interference only affects part of the continuum observing bandwidth. While the location of the interference is not important, the added power in the integrated band in each data record is relevant. The percentage data loss resulting from detrimental narrow band interference may be determined measured from the time series of data records and shall be less than 2\% for a single interferer, as long as the aggregate flux density of the narrow band signal does not exceed the continuum threshold presented in Recommendation ITU-R RA.769. 
Time-variable and frequency-variable interference during continuum observations can be investigated with the above methodology to determine its variability. This variability also includes the effects of Doppler components for the interfering source. The impact of the data loss depends again on the spectral signature of the interference while the location within the band is not important. The percentage of data loss may be measured from the time-series of continuum data records.

\subsection{Spectral line observations}

Spectral-line emissions are often narrow and extremely weak and can be easily overwhelmed partially or totally by an interference signal, making the recovery of the astronomical data impossible. Astronomical spectral line emissions may display significant Doppler broadening because of internal motions in the astronomical source, and often display complex (multi-component) and time-variable emission profiles (see Figure \ref{fig:fig3}). Differentiation between a spectral line of natural origin and an interference signal is often difficult. In general, a spectral analysis is performed using the whole radio astronomy band in order to detect spectral lines at offset radial velocities. Spectral analysis requires calibration of the strength of the spectral feature. For that purpose the baseline must be determined by interpolation of the background spectral power flux density using at least two clean band segments on either side of the spectral line. The accuracy of this procedure depends on the number of interference-free channels in these clean band segments. Therefore, interference at any location within the observing band could render all the spectral line data unusable and must be treated as data loss for all channels.
 
\begin{figure}
\begin{center}
\includegraphics[width= 14cm]{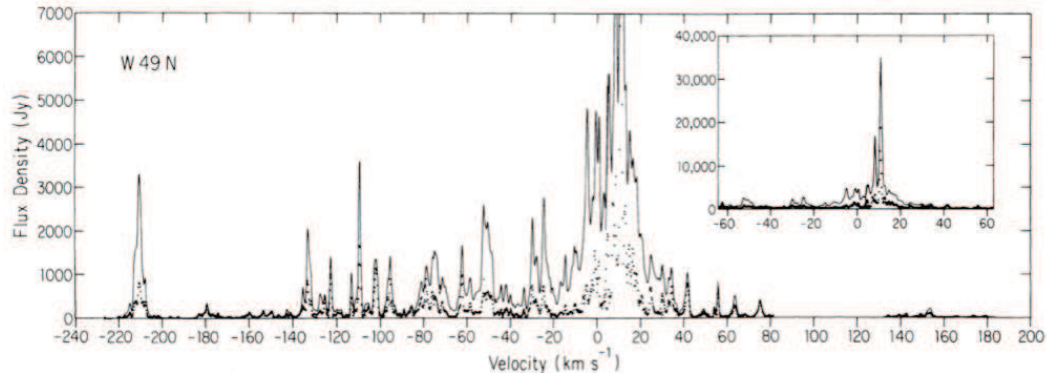}
\end{center}
\caption{Spectral line measurements of the strong H2O maser emission at 22.235 GHz in the Galactic source W49N. Water vapour masers trace the active star formation in as many as 386 structural components of the source. The horizontal axis denotes the radial velocity (VLSR) of the source components in km/s. The velocity resolution is 0.27 km/s or 20 kHz. The velocity extent of the features over the range -230 to +170 km/s in the spectrum covers 37 MHz. The vertical axis denotes antenna temperature in Jansky flux density units. The spectrum has been obtained from interferometry observations by Walker, Matsakis \& Garcia-Baretto (1982).}
\label{fig:fig3}
\end{figure}

{\it \bf Broadband interference} --
Broadband interference covering a large part of the allocated frequency band in excess of the spectral line thresholds in Recommendation ITU-R RA.769 results in a complete loss of a spectral line observation data records. For this case the data loss relates to the percentage of time for lost spectral line data records. 

{\it \bf Narrowband interference} --
An unfavourable location of the interference signal within the observed spectrum may render the spectral line observations not useable for further analysis. In addition, strong interference features may render the complete spectral line spectrum unreliable because of sideband emissions and unwanted effects on the bandpass of the receiver.
Interference limited to one or a few fixed channels may render the data unusable if it falls close to (or is superposed on) an (anticipated or known) astronomical line signal, or if it destroys the clean reference channels on either side of the astronomical spectral line.
For the case of frequency constant narrowband interference during spectral line observations, the percentage data loss resulting from detrimental interference is to be measured in the time-frequency domain.  This procedure does not take account of the actual damage to the spectral line data, as related to location of the interference. Frequency-variable narrowband interference covering a significant part of the RAS bandwidth directly affects the astronomical spectral line signature and cannot easily be removed from the data. This is particularly true for weaker interference with an unpredictable location in the spectrum. In this case, all RFI affected spectral records will be counted as lost data for all spectral channels.
Narrowband interference that sweeps frequency with a cyclic period should be evaluated for continuum data loss using a reference measurement interval longer or equal to this cyclic period. 

\subsection{Time-critical astronomical observations}

Time-variability of astronomical sources is of increasing scientific interest. Modern signal processing makes phenomena with short timescales accessible for radio astronomers but it also reveals their susceptibility to short-timescale interference. The occurrence of interference at critical times during well-timed (and often long-planned) observations may thus jeopardise (or destroy) unique data that is essential for interpreting the astronomical phenomenon. Unrecoverable data loss would thus have a far-reaching effect on the research effort and should be kept at a minimum. 

The protection criteria for this subset of continuum observations that also includes the studies of pulsars and radio transients require further evaluation.  The criteria for these special observations may vary considerably from those given in Recommendations ITU-R RA.769 and ITU-R RA.1513, particularly in the presence of highly time-variable or intermittent interference, and need to be considered for incorporation. 

\subsection{Time-varying propagation conditions}

Time-varying propagation conditions may result in variation of the strength of the interfering signals.  Recommendation ITU-R RA.1031 specifies a percentage of time of 10\% for propagation calculations. This percentage does not automatically lead to a 10\% data loss because propagation conditions vary episodically on timescales of days, which may result in contaminated data for only a few days over periods of weeks. These effects occur primarily at longer wavelengths, i.e. below 1 GHz. 

\section{Conclusions}

The power of interference signals may be determined using measurement intervals commensurate with the time and frequency variability characteristic of the interfering signals.

The detrimental threshold levels for interference during shorter measurement time intervals or different measurement bandwidths may be appropriately scaled from the values presented in Recommendation ITU-R RA.769 for a 2\,000 seconds reference time interval.

The data loss for astronomical instruments may be measured as a percentage of occupancy in the time-frequency domain using the methodology presented in this paper.

In addition to determining the data loss resulting from short-term variability of the interference, primary bands allocated to the RAS may also be evaluated for data loss resulting from the accumulated power flux density of the interference during longer time intervals using the methodology presented in this paper.

The observed time-frequency occupancy characteristics for non-geostationary satellite systems and earth stations in the mobile-satellite service may be incorporated into a effective power flux density simulation to obtain an effective data loss and sky blockage
due to these services.

\end{document}